\documentclass[a4paper,11pt]{article}
\usepackage[colorlinks,allcolors=blue]{hyperref}
\usepackage[T1]{fontenc}
\usepackage[utf8]{inputenc}
\usepackage{lmodern}
\usepackage{graphics}
\usepackage{graphicx}
\usepackage{floatrow}

\title{Higher energy modes of Fractional quantum Hall Effect}
\author{Debashis Das, Moumita Indra, Dwipesh Majumder}

\begin{document}

\maketitle

\begin{abstract}
We have calculated the energy spectra for almost all the filling fraction in the Jain series, low energy fundamental mode as well as higher energy modes using CF theory. The nature of low energy mode and higher energy mode is nearly identical roton mode. We have observed that, these series of filling fractions $\nu=\frac{n}{2pn+1}$ posses $n$ number of roton minima in their higher energy mode.
\end{abstract}

\section{Introduction}

Fractional quantum Hall effect (FQHE) \cite{Tsui82, Laughlin83, Tsui93} in condensed matter physics has unveiled an interesting area to explore collective properties of strongly correlated two dimensional electron system (2DES) in presence of strong magnetic field perpendicular to the two-dimensional (2D) plane, where perturbation theory is practically impossible due to the lack of small parameter in the Hamiltonian within the lowest Landau level(LLL) \cite{LLL}. In addition of 2DES, there is another exciting system of ultra-cold rotating dilute Bose gas, harmonically trapped in 2D; for which there is a possibility of FQHE with those Bose atoms. From the last two decades, theoretically it becomes more interesting research field after the discovery of Bose–Einstein condensation (BEC) \cite{BEC}.  

Interaction between electrons is taken into account by mapping strongly correlated FQH states to the non-interacting integer quantum Hall states (IQHS) with the neutral quasiparticle, called the composite fermion (CF) \cite{Jain_89}. CF's are actually the bound state of electron and even number ($2p$) of magnetic quantum vortices. These composite particles experience a reduced amount of magnetic field,
\begin{eqnarray}
B^* = B - 2p \rho \phi_0 
\label{effB}
\end{eqnarray}

where $B$ is the applied magnetic field perpendicular to the 2D plane, $\rho$ is the density of electron (i.e. of CF), $\phi_0=hc/e$ be the magnetic flux quantum. In this reduced magnetic field CFs form new kind of Landau level (LL), called the $\Lambda$ level. The filling fraction of the $\Lambda$ level is given by $\nu^* = \rho\phi_0/|B^*|$.   
Hence the relation between the filling fraction of electron($\nu$) and the  filling fraction of CF is
\begin{eqnarray}
 \nu = \frac{\nu^*}{2p\nu^*\pm 1}
\end{eqnarray}
where negative sign corresponds to the situation that, as if flux attachment is in opposite direction. 

Most of the observed FQHE in LLL can be explained well by the composite fermion theory, proposed by J.K. Jain.
A particular IQHE state of CF gives a series of FQHE of real electron system with different magnetic flux quanta attachment. Just like, lowest IQHE of CF (i.e. n=1) gives $\nu=1/3$, 1/5, 1/7,$\cdots$ and IQH state with n=2 gives $\nu=2/5$, 2/9, 2/13, $\cdots$ etc, considering positive flux attachment case. These FQH states are known as Jain's sequence. CF theory beautifully explains different partially polarised state of 2-component FQHE, although there are some polarised states beyond CF theory \cite{pola}, which are explained by Chern-simon's approach     \cite{Lopez_Fradkin} and Plasma picture of Halperin \cite{Halperin}.  
When filling fraction of LLs becomes very low, liquid FQH state is destroyed and it is thought that, particles aggregrate to form crystalline structure, called Wigner crystal state \cite{WC, bubbles, strips}.

\begin{figure}
\floatbox[{\capbeside\thisfloatsetup{capbesideposition={right,bottom},capbesidewidth=4.0cm}}]{figure}[\FBwidth]
{\caption{The collective excitation can be expressed as the combination of excitons of CF. The cartoon shows an exciton of $\nu=\frac{3}{6p \pm 1}$ filling fraction, consist of a hole in the 2nd $\Lambda$ level ($n_i$) and a particle (CF) in the 5th $\Lambda$ level ($n_f$). Actual excitation is the combination of many excitons.}\label{exciton}}
{\includegraphics[width=7cm]{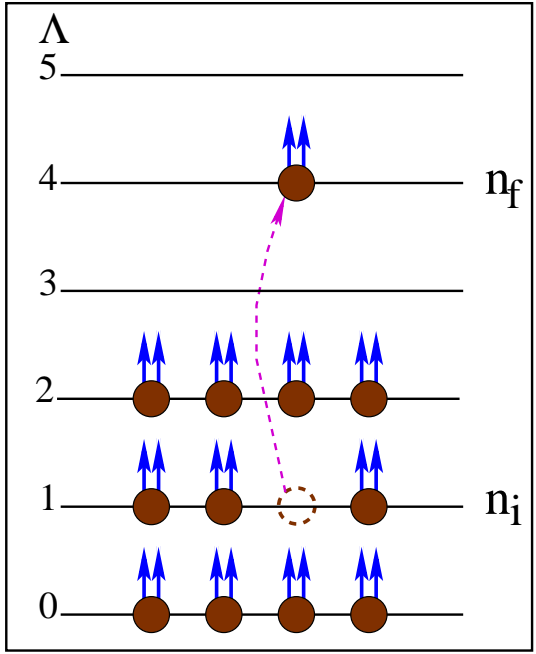}}
\label{exciton}
\end{figure}

Neutral collective modes in quantum Hall states have been studied using the Hartree-Fock approximation \cite{HFA} for the electronic excitons, exact diagonalization \cite{Exact} for small systems, density modulation \cite{DMA} over the ground state in the single-mode approximation (SMA) \cite{SMA}, a Hamiltonian description of composite fermions \cite{HT}, and excitons of composite fermions \cite{CF-Exciton}. All these studies qualitatively and quantitatively describe the presence of “magnetorotons” in the collective modes that have been identified in several inelastic light scattering (ILS) experiments \cite{ILS, DM107, DM106}.
In CF picture the natural collective excitation can be explained as CF-exciton (see the Figure \ref{exciton}), which is a strongly bound state of a CF-hole in a filled $\Lambda$ level and a CF-particle in the empty $\Lambda$-level \cite{Kamilla97, DM14}. There are many studies of higher energy excitation with or without spin \cite{DM107, DM106, DM14, HigSpin}, after the discovery of second mode of excitations \cite{Hirjibehedin05}. The higher energy mode (HEM) is obvious in CF-picture, the excitation of CF across more than one $\Lambda$-levels. 
 
Lowest order collective excitation of almost all the filling fraction in the Jain series of positive flux attachment states has been studied earlier \cite{DDMI}. here, we have presented the results of quantum Monte-Carlo calculation of fundamental mode as well as HEM of excitation for four different series of filling fractions $\nu=1/(2p+1), \; 2/(4p+1), \; 3/(6p+1); 4/(8p+1)$. The HEM of collective excitation of first filling fraction of each series $i.e.$ $\nu=1/3,\; 2/5,\; 3/7$ are there in the literature \cite{DM14}. We have compared the existing results to our results of different filling fraction. We have restricted ourselves in the fully spin-polarized states of every filling fraction, so that kinetic energies of the electrons are frozen and we have only considered Coulomb interactions between the electrons as the Hamiltonian of this system.

\section{Wave function and Numerical technique}

We have figured out spherical geometry \cite{Jain_book, Haldane}  for our numerical calculations instead of plane 2D geometry to avoid edge effect for finite system. It is assumed that, N number of correlated electrons moving around 2D-surface of sphere and a Dirac monopole of strength Q is placed at the centre of the sphere, which gives rise to a radial magnetic field \cite{Haldane} of total flux $2Q\phi_0$ throughout the surface of the sphere. The radius of the sphere can be easily obtained as $R = \sqrt{Q} $, in units of magnetic length $l=\sqrt{\hbar c/eB }$.

The single particle states in this geometry (Landau levels) are nothing but the monopole harmonics, $Y_{n{\bf l} m} (\theta, \phi)$, where $n$ is the LL index, ${\bf l} =Q+n$ is the angular momentum of the particle in the $n$ the LL and $m$ is the z-component of the angular momentum. Obviously the degeneracy of $n$th LL is $2(n+Q-1)+1=2(n+Q)-1$, which depends on the LL on the contrary of the plane geometry. 
 
The effective flux experienced by the CF is given by the equation (\ref{effB}) as $2q=2Q-(N-1)2p$. We choose the value $Q$ in such a way so that the state at $q$ is an integral quantum Hall state at filling $\nu^* = n$.

The ground state wave function of filling fraction $\nu=n/(2pn+1)$ in CF picture can be expressed as projection on LLL of Slater determinant times the Jastrow-factor.

\begin{eqnarray}
\Psi_{\nu=\frac{n}{2pn+1}} =P_{LLL} J \Phi_n
\end{eqnarray}
where $P_{LLL}$ is the projection operator onto the LLL, $\Phi_n$ is the Slater determinant (SD) of $n$ filled $\Lambda$ levels and the Jastrow factor is given by

\begin{eqnarray}
  J = \prod _{i<j}^N (u_i v_j - u_j v_i)^{2p}
\end{eqnarray}
where $u_i=cos(\theta_i/2) \; exp(-i\phi_i /2)$ and $v_i=sin(\theta_i/2) \; exp(i\phi_i /2)$ with ($0\le \theta_i \le \pi$ and $0\le \phi_i \le  2\pi$) is the position of the $i$th particle on the sphere.

\begin{figure}
\floatbox[{\capbeside\thisfloatsetup{capbesideposition={right,bottom},capbesidewidth=4.0cm}}]{figure}[\FBwidth]
{\caption{\small Energy spectrum for two different series of filling fraction $\nu=\frac{1}{2p+1}$ and $\nu=\frac{2}{4p+1}$. The filling fractions have been shown inside the boxes. More than four different number of particles have been considered to produce the result, we have represented the best fit of the calculated results. To get the thermodynamic nature we have used upto 200 particles. The energies are quoted in units of $e^2/\epsilon l$, where $l$ is the magnetic length and $\epsilon $ be the dielectric constant of the sample.  }\label{spec12}}
{\includegraphics[width=7.5cm]{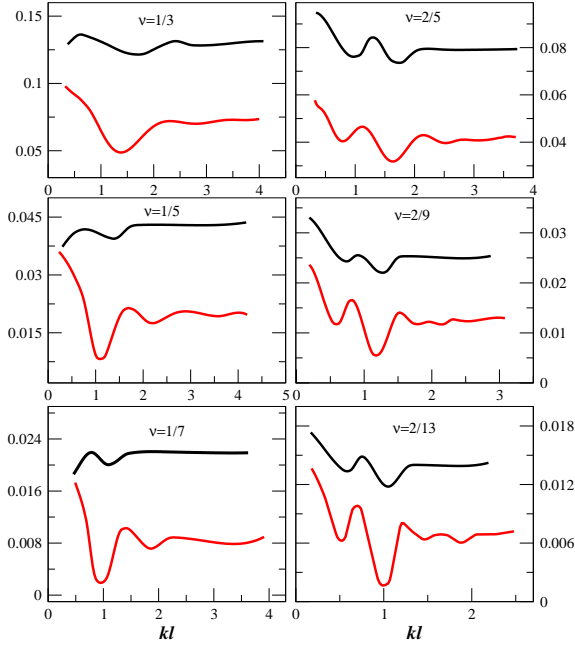}}
\end{figure}

 The excited state wave function of $N$-electron system at the filling fraction $\nu=n/(2pn+ 1)$ corresponding to the transition of a CF from a filled $\Lambda$ level $n_i\;( n_i\le n-1)$ to an empty $\Lambda$ level $n_f\; (n_f\ge n)$, in the spherical geometry is given by \cite{Kamilla97, DM14, Scarola}
\begin{eqnarray}
  \Psi^{n_i\rightarrow n_f }_\nu (L) 
  =P_{LLL} J \sum_{m_h} |n_i m_h; n_f m_p> \; <n_i+q, m_h; n_f+q,m_p|L,0>
\end{eqnarray}
where $|n_i m_h; n_f m_p>$ is the SD of $n$ number of filled $\Lambda$ level with one CF with Z-component of angular momentum $m_h$ has been shifted from $n_i$ (leaving a hole of angular momentum $-m_h$) to $n_f$ $\Lambda$ level with Z-component of angular momentum $m_p$; $<n_i+q, m_h;n_f+q,m_p|L,M>$ are the Clebsch-Gordan coefficients, $L$ is the total angular momentum and M is the corresponding Z-component of total angular momentum. The wave function is very much complicated and time consuming as we need to find the $2(q+n_i)$ number of SDs ($|n_i m_h; n_f m_p>$) in each iteration for a single exciton. 
The SD of an exciton, $|n_i m_h; n_f m_p>$ is basically the ground state with replacement of one single particle state ($n_i m_h$) by state ($n_f m_p$). We calculate the inverse of the ground state matrix to get the cofactor, then we sum the new single particle state times the corresponding cofactor to get the exciton SD.
In our numerical calculation we have considered the sub-Hilbert space  with zero Z-component of total angular momentum ($M=0$) without any loss of generality, to reduce the computation time and resources.

 Actual collective excitation is not a single CF-exciton state rather the superposition of all possible excitons. The excitons are not orthogonal, we have used Gram-Schmidt Orthonormalization procedure to orthogonalize low energy exciton states with a fixed angular momentum. The method of calculation of energy of such kind of mixed state is called CF-diagonalization \cite{CFD}.

The ith excited mode of spectrum with respect to the  ground state  $\Phi_{\nu}$  is given by 
\begin{equation}
  \Delta^i_\nu(L) = \frac{< \chi^i_\nu(L)| H |\chi^i_\nu(L)>}{<\chi^i_\nu(L) |\chi^i_\nu(L)>} - \frac{<\Phi_\nu| H |\Phi_\nu >}{<\Phi_\nu | \Phi_\nu >}
\end{equation}
where \{$\chi^i_\nu(L)$\} are the orthogonal states out of \{$\Psi^{n_i\rightarrow n_f }_\nu (L)$\}.

The Hamiltonian of the system $H=\sum_{i<j} \frac{e^2}{\epsilon r_{i,j}} $, $\epsilon$ being the dielectric constant of the background semiconductor. The multidimensional integration has been carried out using quantum Monte Carlo(MC) method.

\subsection{Finite width correction}

In reality, ideal 2DES doesn't exists. There is a finite width of the 2DES perpendicular to the plane. We have considered this finite width using local density approximation (LDA) \cite{LDA}. 

In our MC calculation we have used the effective 2D potential co consider the finite extension of electron wave function perpendicular to the plane of the sample
\begin{eqnarray}
	V(r_{\bot}) = \int \frac{1}{\sqrt{r_{\bot}^2 + (z_1 - z_2)^2}} |\xi(z_1)|^2 |\xi(z_2)|^2 dz_1\; dz_2
\end{eqnarray}
where $z_1$ and $z_2$ are the Z-component of coordinates of two particles and $r_{\bot}$ is the in plane separation distance between the particles, $\xi(z)$ is the extension of the single particle wave function perpendicular to the plane of an electron, obtained by solving the Schrodinger equation and Poisons equation self-consistently within LDA approximation.

\begin{figure}
\floatbox[{\capbeside\thisfloatsetup{capbesideposition={right,bottom},capbesidewidth=4.0cm}}]{figure}[\FBwidth]
{\caption{\small Energy spectrum for two different series of filling fraction $\nu=\frac{3}{6p+1}$ and $\nu=\frac{4}{8p+1}$ in the fully polarized state. More than four different number of particles have been considered to produce the result, we have represented the best fit of the calculated results. To get the thermodynamic nature we have used upto 200 particles. The filling fractions have been shown inside the boxes. The energies are quoted in units of $e^2/\epsilon l$, where $l$ is the magnetic length and $\epsilon $ be the dielectric constant of the sample. }\label{spec12}}
{\includegraphics[width=7.5cm]{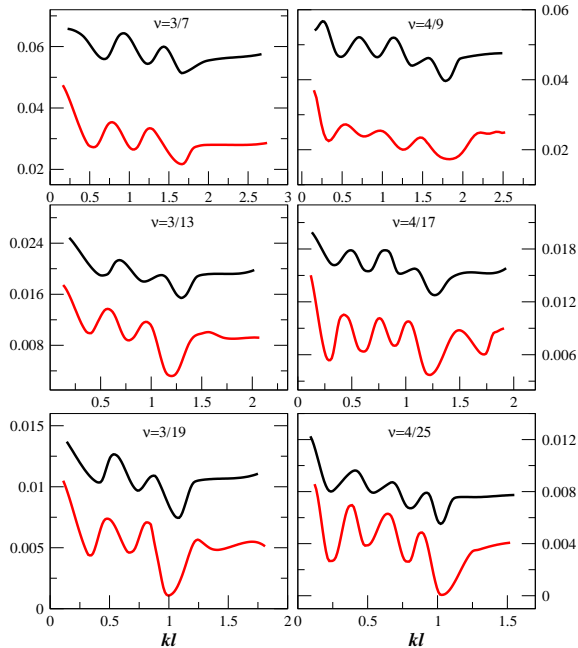}}
\end{figure}

\section{Results and conclusions}

We have calculated the energy spectrum of HEM as well as fundamental mode of excitation for each series of different filling fractions at a particular density of electrons ($n= 10^{11}$ cm$^{-2}$) for a square quantum well of width 30nm. To get the actual excitation energy we have used CF-diagonalization over 5 to 18 number of CF-excitons depending upon the number of filled $\Lambda$ levels.

In the left panel of figure 2, we have plotted the energy spectra for the Laughline (1st series of Jain sequence) filling fraction and in the right panel, we have shown the energy spectra for the second series of Jain sequence. The HEMs of $\nu=1/3$ have been studied previously \cite{DM14}, and here we have included the HEMs of the other filling fractions $\nu=1/5$,1/7 in this series.  

The energy spectra are almost same for 1/3 filling fraction except the value of the energy. We have seen that the value of energy of next HEM for different filling fraction in each series reduces monotonically with the flux attachment but the energy spectrum of each series are identical in nature. It is observed that the position of the roton minimum of HEM moves to the lower value of wave vector for each series with the increase of the flux attachment.
In figure 3 we have also plotted the CF exciton dispersion for the fully spin polarized  state for sequence $\nu=\frac{3}{6p+1}$ and $\nu=\frac{4}{8p+1}$ of  Jain's third and fourth series.

In Ref \cite{DM14} it was predicted that the next HEMs at $\nu=\frac{n}{2n+1}$ will have $n$ rotons, here we also observed that next HEMs of four series i.e. $\frac{1}{2n+1}$, $\frac{2}{4n+1}$, $\frac{3}{6n+1}$, $\frac{4}{8n+1}$ have one, two, three and four number of roton minima. We can not calculate the energy spectrum of next HEMs of other series as the spectra become fluctuating with the number of particles and the spectra is not clear and there is a limitation of flux attachment \cite{bubbles} also. Besides Monte Carlo method fails at the higher $\Lambda$ level filling states, as the projected single particle CF-wavefunction diverge.

\section*{Acknowledgement}

Debashis thanks UGC, India (Sr. No. 2121450734, Ref. No. 21/12/2014 (ii) EU-V) and MI acknowledges DST INSPIRE (Code: IF160850, dated 21/09/2016) for the financial support.

\end{document}